\documentclass[preprint,12pt]{elsarticle}

\usepackage{latexsym,amsfonts}
\usepackage{amsmath,amssymb,graphics,setspace}
\usepackage{amsthm}
\usepackage{amscd} 
\usepackage{MnSymbol}
\usepackage{mathrsfs} 
\usepackage{fontenc}
\newcommand{\mc}{\mathcal}

\usepackage{soul}  
\usepackage{paralist}
\usepackage{color}

\journal{TBA}

\begin{document}

\begin{frontmatter}



\title{On normalization of inconsistency indicators \\ in pairwise comparisons}
\tnotetext[t1]{Alphabetical order implies equal contribution}


\author[au1]{W.W.~Koczkodaj}
\cortext[cor1]{Corresponding author}
\address[au1]{Computer Science, Laurentian University, Sudbury ON P3E 2C6, Canada}
\author[au2]{J.-P.~Magnot}
\address[au2]{Lyc\'ee Jeanne D'arc, rue de grande Bretagne, 63000 Clermont-Ferrand; jean-pierr.magnot@ac-clermont.fr, France}

\author[au3]{J.~Mazurek}
\address[au3]{Department of Informatics and Mathematics, School of Business Administration in Karvina, Silesian University in Opava,  
Univ. Square 1934/3, 73340, Czech Republic; mazurek@opf.slu.cz}
\author[au4]{J.F.~Peters}
\address[au4]{Computational Intelligence Laboratory,
	University of Manitoba, Winnipeg, Manitoba, R3T 5V6, Canada and
	Department of Mathematics, Faculty of Arts and Sciences, Ad\.{i}yaman University, 02040 Ad\.{i}yaman, Turkey}
\author[au5]{H.~Rakhshani}
\address[au5]{Department of Computer Science, Faculty of Mathematics, University of Sistan and Baluchestan, Zahedan, Iran; rakhshani@pgs.usb.ac.ir}
\author[au7]{M.~Soltys} 
\address[au7]{California State University Channel Islands,	Bell Tower West 2265, One University Drive, Camarillo, CA 93012,  USA; msoltys@gmail.com}
\author[au8]{D.~Strza{\l}ka}
\address[au8]{Faculty of Electrical and Computer Engineering, Rzesz\'ow University of Technology, Al. Powsta\'nc\'ow Warszawy, 35-959 Rzesz\'ow, Poland; strzalka@prz.edu.pl}
\author[au9]{J.~Szybowski}
\address[au9]{AGH University of Science and Technology, Faculty of Applied Mathematics, al. Mickiewicza 30, 30-059 Krak\'ow, Poland; szybowsk@agh.edu.pl}
\author[au10]{A.~Tozzi}
\address[au10]{Center for Nonlinear Science, University of North Texas, Denton, Texas 76203, USA}
\begin{abstract}
In this study, we provide mathematical and practice-driven justification for using $[0,1]$ normalization of inconsistency indicators in pairwise comparisons. The need for normalization, as well as problems with the lack of normalization, are presented. A new type of paradox of infinity is described.
\end{abstract}
\end{frontmatter}
Keywords: pairwise comparisons, \sep inconsistency indicator, \sep normalization, \sep decision making. 

\section{Preliminaries}
\label{sec:1}

Consistency in preferences (by pairwise comparisons) can be traced to \cite{KS1939}, published by Kendall and Smith in 1939. By 1976, there were four articles listed in \cite{DF1976}, with the inconsistency in the title. In \cite{Saaty1977}, the consistency index (CI) was introduced. Many other definitions followed. All of them attempt to answer the most important question: how far have we departed from the consistent state which is well established by the consistency condition illustrated in Fig.~\ref{fig:fig2} in this text. Briefly, every cycle of three interrelated comparisons: $A/B$, $B/C$, and $A/C$ must be reducible to two comparisons, since the third comparison is a result of multiplication or division of these two comparisons. For example, given $A/B$ and $B/C$, $A/C$ should be equal to $(A/B)*(B/C)$. If $A/B$ and $A/C$ are given, $B/C$ is equal to $(A/C) / (A/B)$. If $B/C$  and $A/C$ are given, $A/B$ is equal to  $(A/C)/ (B/C)$. No other combination exists for a cycle of three comparisons. If all three ratios are given, inconsistency in such cycle may take place and usually, it does. Earlier definitions of inconsistency were based on the total count of inconsistent triads. Evidently, such cardinal inconsistency was imprecise.

This study proposes to regard the inconsistency in pairwise as a degree or extent of disagreement modeled on the concept of probability measure. Our study also demonstrates a problem shown in \cite{KSW}, where the normalization was not used.

\section{Inconsistency indicator as a degree of disagreement in pairwise comparisons}

The interval $[0,1]$ (or discrete values between 0 and 1) is used in many theories and situations including:

\begin{itemize}
	\item probability,
	\item multi-valued logic (as proposed by Lukasiewicz), 
	\item fuzzy logic, 
	\item rough set theory, 
	\item theory of evidence. 
\end{itemize}
\label{use01}

There are also a number of zero-one laws of which the convergence of conditional expectations, known as ``L\'evy's zero–-one law,'' can be summarized as ``if we are learning gradually all the information that determines the outcome of an event, then we will become gradually certain what the outcome will be,'' which has application to the consistency-driven pairwise comparisons (recently used in \cite{KKL2014}, but based on inconsistency indicator introduced in \cite{K1993}).
The important interpretation of ``L\'evy's zero–-one law'' is ``the gradual data gathering to determine the outcome of an event, then we will become gradually certain what the outcome will be'' since it has the direct application to the reduction of inconsistency in pairwise comparisons

According to \cite{MathEncycl}, P. L\'evy proved in 1937 (see \cite{Levy1937}), that Kolmogorov's theorem follows from a more general property of conditional probabilities, which says that \\
\begin{equation}
\label{eq:1}
\lim_{n \rightarrow \infty} P\{f(X_1, X_2, \ldots, X_n) |(X_1, X_2, \ldots, X_n  )\},
\end{equation}

almost certainly equals 0 or 1 (depending on whether $f(X_1, X_2, \ldots X_n)$ is zero or not). 

The use of interval $[0,1]$ for inconsistency indicators is not only coming from the Occam's razor principle. The Section~\ref{sec:3} shows that the lack of normalization in \cite{KSW} leads to problems which this study attempts to correct. In this study, we will follow a business principle ``change a problem into opportunity'' and use it not only to correct \cite{KSW} but to reason for an important axiom.

\section{The need for normalization of inconsistency indicators}
\label{sec:3}

\noindent The fallacy of the Definition 3.6 on page 83 in \cite{KSW}: 

\begin{quotation}
Let $X=\langle X,\cdot\rangle$ be a group. 
A $\mc{G}$-distance-based inconsistency indicator map (in abbreviation: a $\mc{G}$- inconsistency indicator map) on the group  $X$ is a function $T: X^3\to G$ such that, for all $a,b,c,d,e\in X$, the following conditions are satisfied:
\begin{enumerate}
	\item[(i)] $T(a,b,c)=1_G\Leftrightarrow ac=b$;
	\item[(ii)] $T(a,b,c)=T(b, ac, 1)$;
	\item[(iii)] $T(a, de,c)\leq T(a,b,c)\odot T(d, b, e)$.
\end{enumerate}
\end{quotation}

\noindent leads to a contradiction in the theory presented there.

The fallacy in the above definition is evidenced below by using the sequence of triads $T_n$ converging to a consistent matrix when the limit of $n$, a natural number, is the infinity.

Formally, an inconsistency indicator ($ii$) is defined for a PC matrix, say M. So, $ii(M)$ should be used.
However, a PC matrix:

$$M= \left[ 
\begin{array}{ccc} 
1 & x & y \\ 
1/x & 1 & z \\ 
1/y & 1/z & 1
\end{array} \right] $$

is reduced to the three values: $x,y,z$ which generate the entire matrix. For this reason, 
we will be using $ii(x,y,z)$ as a notation shortcut (to save on typing and space).

To make our point, all we need to do is demonstrate just one counter-example
(for any particular distance). In particular, our goal is achieved when we construct all triads
in the sequence $T_n$ this way so that they have constant inconsistency greater than 0 (e.g., one or $2^{64}$) and yet such sequence converges. In fact, it is enough to show the convergence for any sequence (e.g., Cauchy sequence). Evidently, $T_n$ is a Cauchy sequence. 
The well-established consistency condition for PC matrix $M=[m_{ij}]$ is:
\begin{equation}
\label{eq:5}
m_{ik}=m_{ij}*m_{jk}~~~ \mbox{for~}i<j<k\leq n
\end{equation}

These three PC matrix elements create a triad $(m_{ij},m_{ik},m_{jk})$ denoted here as $(x,y,z)$. A triad is consistent when its middle value is equal to the multiplication of the first and third value. To show that Definition 3.6 in \cite{KSW} leads to contradiction, consider the following sequence of triads:

\begin{equation}
\label{eq:2}
T_n=(x^n, x^{2n}+1, x^n)
~~~ \mbox{~~~for~natural}~ n=1,2, \ldots,  \mbox{~and for~} x>1. 
\end{equation}

\noindent The Euclidean distance between the middle value in the triad $T_n$ and the result of two other values after the group operation (multiplication) is constant and equals to:  $d(x^{2n}+1, x^n\cdot x^n)=1$.  The relative error is defined as:

\begin{equation}
\label{eq:3}
\eta = \left|  \frac{y-y_{approx}}{y}  \right|.
\end{equation}

When the relative error definition, given by the equation~(\ref{eq:3}), is applied to the middle value of the triad $T_n$, we get:

\begin{equation}
\label{eq:4}
\frac{d(x^{2n}+1,  x^n\cdot x^n)}{x^{2n}+1}= \frac{1}{x^{2n}+1} \rightarrow 0 ~~~ \mbox{~~~for} ~ n \rightarrow \infty.
\end{equation}


\begin{figure}
	\centering
	\includegraphics[width=0.7\linewidth]{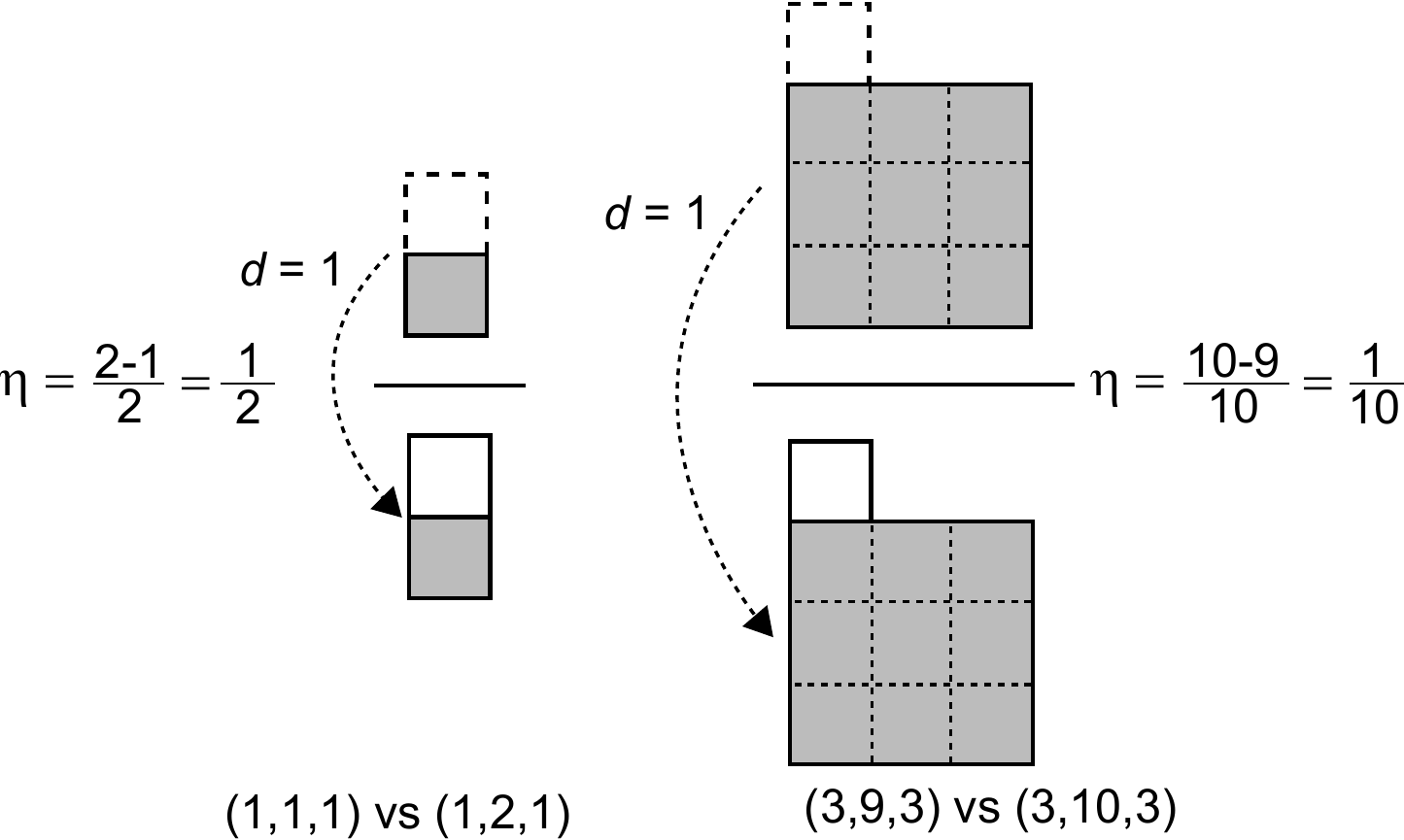}
	\caption{Example of two triads with identical $d$}
	\label{fig:fig1}
\end{figure}

To illustrate that the distance (generalized or not) cannot serve as an inconsistency indicator, all we  need to do is to produce just one counter-example. Fig.~\ref{fig:fig1} shows that ``one extra square'' added to a consistent triad $(1,1,1)$ (hence ``one of two'') is worse than one extra square added to nine squares.

\noindent It is worthwhile to notice that the following triad: 
\begin{center}
	$(\infty, \infty^2, \infty)$ 
\end{center}
is simply reduced to: 
\begin{center}
	$(\infty, \infty, \infty)$
\end{center}
\bigskip
So, a triad may look like $(x,x,x)$ but $(x,x,x)$ is only consistent for $x=1$, while $(x,x^2,x)$ if consistent for any $x$ (technically, $x > 0$ since PC matrix cannot contain zero or negative values). 


From the point of view of the proof of convergence of $T_n$, it does not matter how the real limit looks like: $(x,x^2,x)$ or $(x,x,x)$ as long as the consistency condition holds for it.

Replacing 1 in $T_n$ with any constant (e.g., $2^{64}$ which is a cosmic number) just increases $n$ but the entire reasoning is unchanged. For the disbeliever in $\infty$, we can use three equal sticks of the length 1 which ratios were inaccurately estimated to $(1,2,1)$ giving the relative error of $|2-1|/1$, hence, 100\%. Three other sticks, each 10 times shorter than the other stick, with the ratio estimations as $(10,101,10)$ give the 1\% relative error while the Euclidean distance is in both cases is identical and equal to 1.
The sequence of triads $T_n$ converges to a consistent triad although each of its element has a constant inconsistency indicator making it inconsistent.
Defining the inconsistency indicator as an unnormalized distance leads to the following contradiction: 
\begin{quote}
	a sequence of triads converges to a consistent triad, however, each triad has (according to the theory in \cite{KSW}) the inconsistency indicator is equal to 1 for each triad in the sequence; therefore, 1 has to be the sequence limit.
\end{quote}

The fallacy in the definition of inconsistency in pairwise comparisons as distance for the multiplicative case in \cite{KSW} extends to the additive case as Fig.~\ref{fig:diagram1} illustrates it. Areas of triangles are proportional to ratios they represent: 2, 5, and 3. The layout of these values resemble their location in a PC matrix given as:

\begin{displaymath}
M = \begin{bmatrix}
1 & A & B \\
\frac{1}{A} & 1 & C \\
\frac{1}{B} & \frac{1}{C} & 1
\end{bmatrix}
\end{displaymath}

The first value in a triad $(x,y,z)$ represents a ratio of entities $A$ and $B$: $x=A/B$. The middle triad value $y=A/C$ is located in the same matrix row as $x$ to the right of $x$. The third value in the triad $z=B/C$ is located in the same column as $y$ below it. Values above the arrows represent the value for the third node to make the entire triad consistent. For example, if the value of 2.5 above the upper arrow is used for the replacement of value 3 in the third node (which this arrow does not connect), the triad becomes consistent (since 2.5*2=5).
Similarly, using the value of 6 above the arrow connecting $x$ and $y$ for $z=A/C=5$ makes the triad $(x,y,z)$ consistent.
    
\begin{figure}[!ht]
	\centering
	\includegraphics[width=0.7\linewidth]{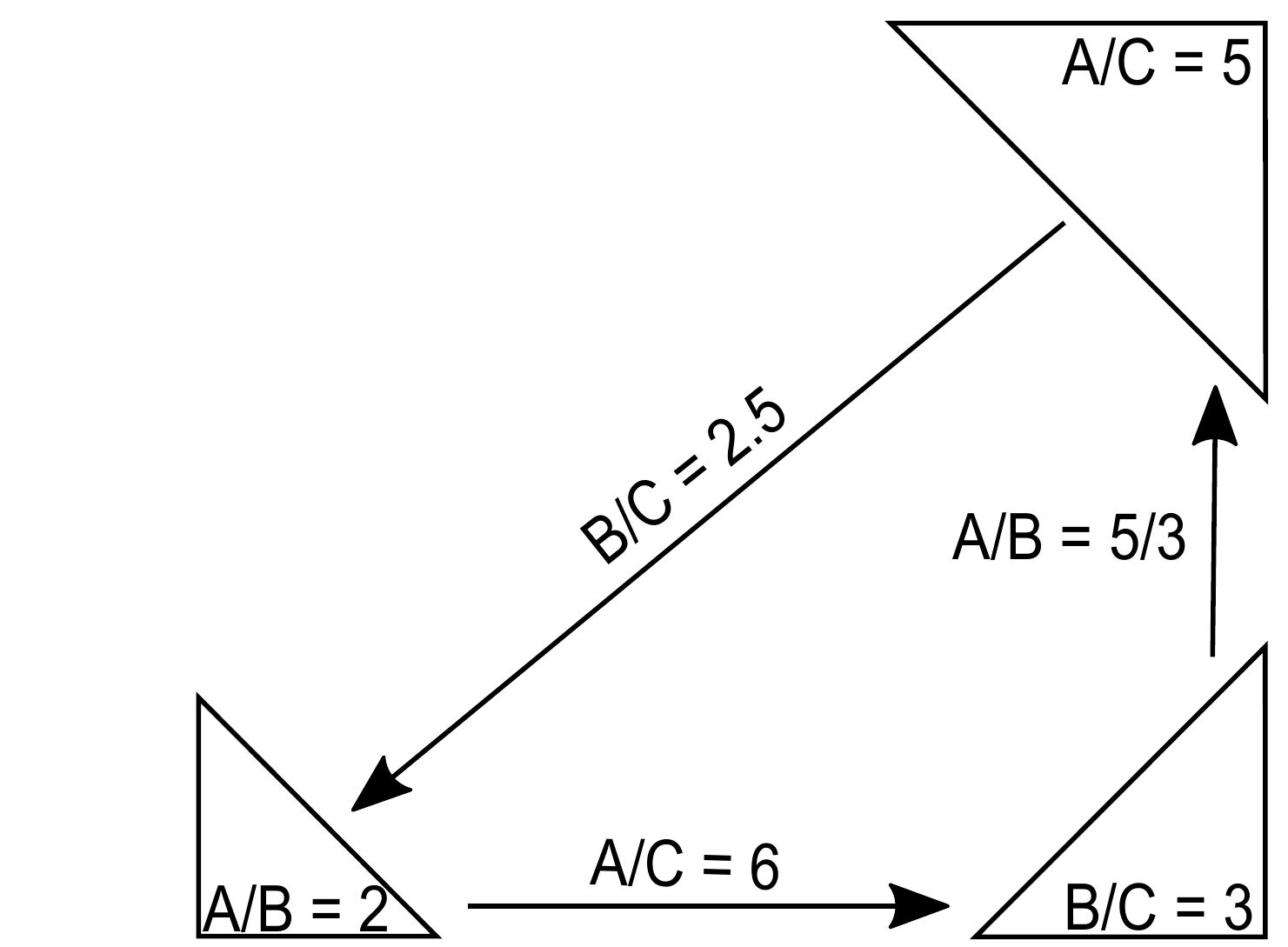}
	\caption[inconsistency concept]{Illustration of inconsistency concept}
	\label{fig:fig2}
\end{figure}
 
The above reasoning demonstrates why using the distance, which is not normalized, as inconsistency is unacceptable. The distance between $y$ and $xz$ in a triad $(x,y,z)$  remains constant (e.g. 1 or $2^{64}$) for all triads in the sequence $T_n$ while the limit of the relative error of approximation is converging to 0, hence, the triad becomes consistent at its limit with the expected inconsistency 0.

It is worthwhile to notice that the relative (hence, normalized) error was needed to make the point but it cannot be taken as a proof of the sufficiency. The analogy to probability provides enough reasoning for the sufficiency.

The above reasoning applies to \cite{CD}. In particular, \cite{CD} includes:
\begin{quote}
	 ``DEFINITION 6.1. The consistency index of the matrix in Equation 6.1 is given by $I_\mc{G}(A) = ||\rho_{123}|| = d_\mc{G}(a_{13}, a_{12} \odot a_{23}).$ (6.2)''
\end{quote} 
\noindent which was analyzed above.

The similar reasoning applies to the additive case of pairwise comparisons, where ratio (``how many times?'')
is replaced by difference (``by how much?'') against the use of the unnormalized distance as an inconsistency indicator.
Fig.~\ref{fig:diagram1} illustrates the interpretation the inconsistency indicator as the area of the ``triangle'' which is reduced to the line for the consistent case.
\begin{figure}[p]
	\centering
	\includegraphics[width=1\linewidth]{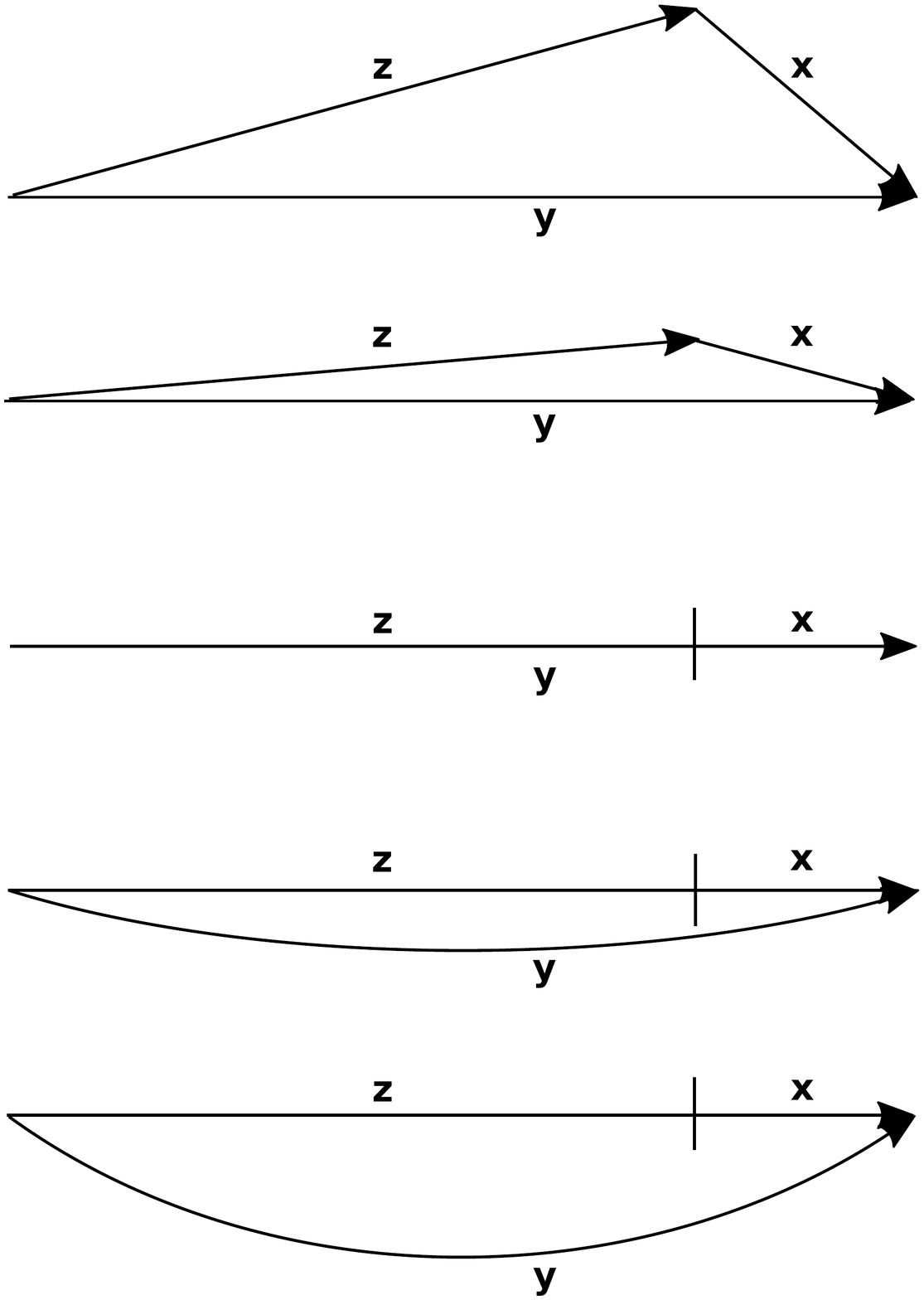}
	\caption[Inconsistency in additive pairwise comparisons]{Inconsistency in additive pairwise comparisons}
	\label{fig:diagram1}
\end{figure}
It is easy to see that the inconsistency for a triad $(x,y,z)$, defined in \cite{K1993} and finally simplified to $ii=1-\min(y/(x*z), x*z/y)$ in \cite{KSz}, does not suffer the above problem since the distance $d(y-x*z)$ is normalized and converges to 0. It is worth mentioning that this may be an example for the need of normalization for applications since the distance normalization seems to be indispensable for the inconsistency indicator definition. The need for a normalized rating scale for pairwise comparisons has been shown in \cite{K2016}. The lack of the rating scale normalization leads to a paradox. Similarly, the use of ``central tendency'' in the inconsistency indicator definition leads to deviations as demonstrated in \cite{KSz} (Section 5 and 6). The two counter-examples and mathematical reasoning were provided that eigenvalue-based inconsistency, used in some customizations of pairwise comparisons, has no mathematical validity since it tolerates arbitrarily large errors when the size of a PC matrix increases to the infinity. 

Our reasoning cannot be interpreted as ``normalization will fix it all'' since not every normalized distance can be used as inconsistency indicator, but whatever inconsistency indicator is defined, it has to be normalized for the reasons listed in the Section~\ref{sec:4}. Evidently, the list of reasons is, in all likeliness, incomplete, but sufficient to normalize all inconsistency indicators.

There is no question that we need to make risky moves in the absence of hard evidence and ``stipulative'' (more relevant term: ``speculative'') definitions may be proposed. However, such definitions should be clearly labeled so other researchers could be alerted and look for a validation or fallacy of such definitions. The theory, based on stipulative foundations, which are not properly labeled, becomes hard to invalidate as time passes and the progress of science suffers. One may only imagine that the eignevalue-based inconsistency in \cite{Saaty1977} was ``stipulative'' since it was invalided in \cite{KSz}) in 2014.

Definition 3.6 in \cite{KSW} could also be regarded as a stipulative definition. However, it was not labeled (or otherwise indicated) as ``stipulative''in the text . The assumption that a generalized distance can be used as a map for inconsistency indicators was arbitrary, not rooted in the deep knowledge of pairwise comparisons, and unverified by real life examples.  Not only is it corrected by this study, but has facilitated the discovery of the need for normalization which is a paradigm shift in the research of the inconsistency in pairwise comparisons.  

It is worth to notice that all coefficients of $T_n$ have the limit $+\infty$ hence we may wonder whether or not such limit is attainable.
Observe that infinity is not a number and was rejected by the intuitionistic mathematics that was advocated by L.E.J. Brouwer. As it turns, the existence of infinity cannot be proved and it needs to be postulated (something that many of us often overlook to do).

One of the philosophical issues of applied mathematics is the approximation theory of real numbers. Most transcendental real numbers
cannot be described since the set of transcendental real numbers  is not countable, while a picture of one number after another can only show a countable set of numbers.
Evidently, at various degrees of knowledge and applications, the idealized framework of mathematics may be questioned by the the common sense with regards to the lack of effectiveness and/or feasibility of obtained results.

\section{Why not $[0,1]$?}
\label{sec:4}

The normalization has become so popular (especially in the applied science) that it's impossible to trace its origins. However, the probability is the high suspect, since it was already used (in its embryonic and partially mystic way) by Cardano, who was a gambler, in the 14th century. The probability has involved into one of the most powerful scientific methods universally used in most branches of applied and many theoretical sciences. 
The normalization in probability theory is related to two of three axioms proposed by Kolmogorov in 1933 hence it is of a considerable importance. However, normalization might have been in use for a few centuries. According to \cite{SV2003, SV2006}, the acceptance of Kolmogorov's axiomatization was not instantaneous but took tens of years (some researchers still try to relax some of his axioms).

The issue of normalization leads to the non-trivial mathematical theory of boundedness and bornology in \cite{Hogbe-Nlend1977bornologies} and in \cite[\S 1.18]{Peters2016computationalProximity} 
is related to compactification.  A mathematical object $A$ such as a set or function is \emph{bounded}, provided that $A$ has a value such that all members of the set or function are less than this bound.  For example, the set of points in the half closed, half open interval $[0,1)$ are bounded above by 1.  Again, for example, the function $f(x) = \log(x), 0\leq x\leq 2$ is bounded above by 1.

A \emph{bornology} on a nonempty set $X$ is a collection $\mathcal{B}$ of bounded sets $B$ that cover $X$ such that $X = \bigcup B$, and $\mathcal{B}$ is stable under inclusions and finite unions.  The power set of a set $X$ (denoted by $(X,2^X$) is the collection of the subsets of $X$, including $X$.  The power set $2^X$ for $\bigcup \{x\}, x \in X$ 
is an example of a bornology.  Boundedness and bornology are closely related to compactification.  The bordification is related to compatification. 

A subset $A$ of a topological space $X$ is \emph{compact}, provided, for every open cover of $A$, there is a finite subcover of $A$~\cite[\S 11.0]{Peters2016computationalProximity}.
For example, every closed and bounded subset in $\mathbb{R}^n, n\geq 2$ is compact.  However, $\mathbb{R}^1$ as well as any interval of $\mathbb{R}^1$ is not compact.  For instance, let $X = [0,1]$.  Then $2^X$ is a topological space that is not compact, since $2^X$ has no finite subcover.  Again, the set of all subsets of a geodetic line the plane is not compact for the same reason. Let $X$ be a non-compact space and let $p$ be a point that does not lie in $X$. The next step is to take a non-compact space $X$ and modify it to make it compact.  The result is a \emph{compactification} of $X$ (see, {\em e.g.},~\cite[\S 2.5]{Krantz2010compactification}).

One of the  most important properties of normalization to $[0,1]$ is the stability under multiplication defined as: 

$$\mbox{for~ }\forall t,u \in [0,1], t\cdot u \in [0,1]$$

The maximality of $[0,1]$ is evident in terms of the upper bound, since we cannot extend 0 to a negative value.
If we extend 1 by the smallest positive value imaginable, say $\alpha >0$, the multiplication gives:
\begin{equation}\label{eq:6}
(1+\alpha) \cdot (1+\alpha) = 1 + 2\alpha + \alpha^2 > 1+\alpha.
\end{equation}


It is worthwhile to notice that every non generative subinterval of $[0,1]$ is also stable under multiplication. However, it is not maximal.
In other words, $[0,1]$ is the only bounded (by two numbers excluding $\infty$) interval of the maximal size containing the multiplication of two non negative real numbers belonging to it.

The ``Zero One Infinity Rule" is attributed to Willem Louis van der Poel (a pioneering Dutch computer scientist, who is known for designing the ZEBRA computer). By the above elimination of $\infty$, our choice for the range of inconsistency indicators is reduced to 0 and 1.


Acting in this spirit of the probability axioms, we postulate the lower and upper bounds for inconsistency indicator as: $ 0 \leq ii \leq 1$ without the loss of generality.
Unlike probability, inconsistency indicators do not need to reach the upper bound value. It reflects the reality that no one can be indefinitely inconsistent with some exceptions such as ``0-1'' inconsistency indicator with 0 for all consistent triads and 1 otherwise. Evidently, $ii=1-min(y/(xz), xz/y)$ for $(x,y,z)$ cannot ever have a value of 1 since neither $xz$ nor $y$ can be 0 ($x,y,z$ are ratios hence must be strictly positive).

Traditionally, most measures are  constructed by setting two thresholds, namely, lower and upper bound. The lower bound is nearly always (if possible) assumed as 0. The upper bound is $10^c$, where $c = 0$ for probability, fuzzy sets, rough sets, and more; $c = 1$ for various indicators such as school grades; $c = 2$ for  percent values (hence approximation error and business).

There are a number of reasons why the interval $[0,\infty)$ is not as good as $ [0,1]$ but the most important reason is that the first half of $[0,\infty)$ is still $[0,\infty)$. In practice, the inconsistency in pairwise comparisons is unavoidable unless we decide to use the simplified version of pairwise comparisons as described in \cite{KS} with the minimum of $n-1$ comparisons of so-called principal generators which allow us to compute the rest of PC matrix elements from the consistency condition ($a_{ij}*a_{jk}=a_{ik}$). When inconsistency exists (and it is nearly always as previously noticed), its degree is of great importance for the level of tolerance similar to $p-$value in statistics. The $p-$value interpretation is the probability that, using a given statistical model, the statistical summary is less than or equal to the chosen significance level ($\alpha$) than the actual observed results (e.g., the difference between mean values of two compared groups). Needless to say that the value $\alpha$ in statistics is arbitrarily assumed for applications. In medical research, $\alpha=0.05$ is frequently assumed as a reasonable significance level. 

In pairwise comparisons, a similar level ``significance'' for the inconsistency. It should be called as level of tolerance.  Using sn interval $[0,1]$ is 
more convenient than $[0,\infty)$. For some inconsistency indicators (e.g., introduced in \cite{K1993}, called $Kii$ here), the level of tolerance is set to 1/3 since $Kii \in [0,1)$, and it does not have easy interpretation in $[0,\infty)$. The related concepts in statistics are confidence levels and confidence intervals  introduced by Neyman in 1937 in \cite{Neyman1937}. For the inconsistency in pairwise comparisons, the issue of a tolerance level is of as fundamental importance as the statistical significance in statistics.

The lack of ``point of reference'' in $[0,\infty)$ was a problem in \cite{Saaty1977}, where a random inconsistency measure was introduced. It is in the direct contradiction of Stevens' observations in \cite{SSS1975} stating that: \begin{quote}
	``the assignment can be any consistent rule. The only rule not allowed would be random assignment, for randomness amounts in effect to a nonrule''.
\end{quote} 

The interval $[0,1]$ is used for most indicators and measures in applied sciences as well as in theoretical sciences. 
Keeping the values in the interval $[0,1]$ may be the most natural way human brains process/assess concepts of nothing $(0)$, a part of $(0,1)$ and total/sum (which is 1). However, the most important reasoning for us is the following celebrated mathematical theorem.

Normalizing mappings is well researched and there are many functions mapping $\mc{R^+}$ onto $(0,1)$. For example,  generalized logistic function or curve, also known as Richards' curve, a Gompertz curve or Gompertz function, named after Benjamin Gompertz, a sigmoid function, and many other functions are normalized.

It is worth mentioning that the inconsistency indicator, proposed in \cite{K1993} in 1993, is normalized. Recently, it was simplified in \cite{KS} to:   

\begin{equation}\label{eq:7}
Kii=1-\min\left(y/(x*z), x*z/y\right) \mbox{~for~} (x,y,z).
\end{equation}

The equation (\ref{eq:7}) has an equivalent form 
\begin{equation}\label{eq:8}
Kii = 1-e^{-|ln(\frac{y}{x\cdot })|}.
\end{equation}
\\
A different (and a bit unusual) type of normalization is needed to prevent a rating scale paradox (addressed in \cite{K2016}). \\

The original title of this section was ``Why $[0,1]$?'' but, after intensive study, it was changed to reflect the reality in the applied science. 
It is mostly pure mathematics and theoretical physics where $\infty$ reigns although the speed limit is observed even in the theoretical physics. Using $\infty$ in proofs is not only highly desirable but often necessary. However, variables of the indicators type are easier to process when they are normalized. For this reason, using [0,1] should be a rule. Such a rule should only be relaxed if there is a good reason not to follow it. 

\section{Conclusions}
\label{sec:5}

The best way is to avoid inconsistency as much as possible. Sometimes, it is possible as \cite{KS} demonstrates. A decision maker may choose to provide just $(n-1)$ pairwise comparisons, which must be ``principal generators'' (e.g., located just above or below the main diagonal and other strategic places specified in \cite{KS}). The remaining PC matrix elements can be reconstructed so the entire PC matrix  is consistent. Using more than $(n-1)$ pairwise comparisons may (and usually does) lead to inconsistency, which is hard to manage without the use of heuristics. 

However, avoidance of inconsistency has a reflection in the use of artificial levees in the river management. When the artificial levee is breached, the damage to properties and lives can be greater than flooding which occurs without an artificial levee constructed.
When we enter more data and properly analyze their inconsistency to reduce it, the expectations are that the end result is improved. We hope that the $[0,1]$ normalization of inconsistency indicators in pairwise comparisons may become one of ``the best practices'' of future knowledge management standards.

In conclusion, we have learned an important lesson here: when we use applied mathematics, we no longer have a complete freedom of using any term in any way we wish. The mathematical distance cannot serve as the inconsistence indicator as we demonstrated here by the use of the relative error.


\section{Acknowledgments}

The research has been supported in part by: 
\begin{itemize}
\item the Polish Ministry of Science and Higher Education 
\item the Ministry of Education, Youth and Sports Czech Republic within the Institutional Support for Long-term Development of a Research Organization in 2017.
\end{itemize}

\section*{References}

\bibliographystyle{9}

\begin{thebibliography}{00}
	

\bibitem{CD} 
B. Cavallo, L. D'Apuzzo,  A general unified framework for pairwise comparisons matrices in multicriterial methods, Int. J. Intelligent Systems, 24: 377--398, 2009.

\bibitem{DF1976}
Davidson, R.R; Farquhar P H, Bibliography on the Method of Paired Comparisons,
Biometrics, 32(2):241-252, 1976

\bibitem{AF1999}	
Jean-Pierre Aubin;  H\'el\`ene Frankowska,
	Set-Valued Analysis (Modern Birkhäuser Classics) 1st ed. 1990. 2nd ed. 2008.
	
%
%



\bibitem{KS1939}
Kendall, M.G.; Smith, B.B., On the Method of Paired Comparisons, Biometrika, 31(3/4):324-345, 1939(1940)



\bibitem{Hogbe-Nlend1977bornologies} 
Hogbe-Nlend, H. ., Bornologies and Functional Analysis.  North-Holland Pub. Co., Amsterdam, 1977, MR 0500064

\bibitem{K1993} 
Koczkodaj, W.W., A new definition of consistency of pairwise comparisons, Mathematical and Computer Modelling,  18(7): 79--84, 1993.

\bibitem{K2016}
Koczkodaj, W.W.,
Pairwise Comparisons Rating Scale Paradox,
Transactions on Computational Collective Intelligence XXII, 9655:1--9, 2016.

\bibitem{KKL2014}
Koczkodaj, W. Kulakowski, Ligeza, A., On the quality evaluation of scientific entities in Poland supported by consistency-driven
pairwise comparisons method, Scientometrics, 99:911–926, 2014.



\bibitem{KSz}
Koczkodaj, W.W.; Szwarc, R. On axiomatization of  inconsistency indicators for pairwise comparisons, Fund. Inform. \textbf{132} (2014), no.4, 485--500.

\bibitem{KS}  
Koczkodaj, W.W., Szybowski, J. Pairwise comparisons simplified, Applied Mathematics and Computation, 253:  387--394, 2015.

\bibitem{KSW}  
Koczkodaj, W.W.; Szybowski, J.; Wajch, E., Inconsistency indicator maps on groups for pairwise comparisons, International Journal of Approximate Reasoning, 69: 81-90, 2016.

\bibitem{Krantz2010compactification} Krantz, S.G., Essentials of Topology with Applications.  CRC Press, London, UK, 2016. 

\bibitem{Levy1937}
L\'evy, P., Th\'eorie de l'addition des variables al\'eatoires, Gauthier-Villars, 1937.

\bibitem{Peters2016computationalProximity} 
Peters, J.F., Computational Proximity.  Excursions in the Topology of Digital Images, Springer, Berlin, 2016.

\bibitem{Saaty1977}
Saaty,  A scaling method for priorities in hierarchical structures. Journal of Mathematical Psychology, 15(3), 234--281, 1977

\bibitem{Neyman1937}
J. Neyman, Outline of a Theory of Statistical Estimation Based on the Classical Theory of Probability,
Philosophical Transactions of the Royal Society of London. Series A, Mathematical and Physical Sciences
Vol. 236, No. 767 (Aug. 30, 1937), pp. 333-380 

\bibitem{S1961}
Slater, P., Inconsistencies in a Schedule of Paired Comparisons,
Biometrika,  48(3-4): 303--312, 1961 

\bibitem{SV2003}
Shafer, G.; Vovk, V., The origins and legacy of Kolmogorov's
Grundbegriffe, Working Paper \#4, 
First posted ``www.probabilityandnance.com'',
February 8, 2003. 
Last revised April 5, 2013.

\bibitem{SV2006}
Shafer, G.; Vovk, V., The sources of Kolmogorov's Grundbegriffe, Statistical science, 21(1):70--98, 2006. 

\bibitem{SSS1975}
Stevens, S.S.. Psychophysics. New York: Wiley, 1975.

\bibitem{MathEncycl}
Zero-one law. Encyclopedia of Mathematics, \\
www.encyclopediaofmath.org, accessed 2017-01-27 

\end{thebibliography}

\end{document}